\documentclass{acm_proc_article-sp}
\usepackage{algorithmic}
\usepackage{algorithm}
\begin{document}

\title{Extending Karger's randomized min-cut Algorithm for a Synchronous Distributed setting}
%
%
%
%
%

\numberofauthors{8} 
%
\author{
%
%
\alignauthor
Shine S \\
       \affaddr{Dept. of Computer Science and Engineering}\\
       \affaddr{College of Engineering Trivandrum}\\
       \affaddr{Kerala, India}\\
       \email{shine@cs.cet.ac.in}
\alignauthor
K. Murali Krishnan \\
       \affaddr{Dept. of Computer Science and Engineering}\\
       \affaddr{National Institute of Technology Calicut}\\
       \affaddr{Kerala, India}\\
       \email{kmurali@nitc.ac.in}
}

\maketitle
\begin{abstract}
A min-cut that seperates vertices $s$ and $t$ in a network is an edge
set of minimum weight whose removal will disconnect $s$ and $t$.  This
problem is the dual of the well known $s-t$ max-flow problem. Several
algorithms for the min-cut problem are based on max-flow computation
although the fastest known min-cut algorithms are not flow based. The well known Karger's randomized algorithm for
min-cut is a non-flow based method for solving the
(global) min-cut problem of finding the min
$s-t$ cut over all pair of vertices $s,t$ in a weighted undirected
graph. This paper presents an adaptation of Karger's algorithm for a synchronous distributed setting
where each node is allowed to perform only local computations.  The paper essentially
addresses the technicalities involved in circumventing the limitations
imposed by a distributed setting to the working of Karger's algorithm.
While the correctness proof follows directly from Karger's algorithm,
the complexity analysis differs significantly.
The algorithm achieves the same probability of success as the original
algorithm with $O(mn^{2})$ message complexity and $O(n^{2})$ time
complexity, where $n$ and $m$ denote the number of vertices and edges
in the graph.
\end{abstract}

\category{}{Distributed Algorithms}{Metrics}[complexity measures]
\category{}{Graph Theory}{Miscellaneous}

\terms{Network-flow}


\section{Introduction}
The problem of computing the minimum-cut in a weighted graph has been
classically studied in literature
as the dual of the well known max-flow problem for networks
\cite{NRMF97} and classical solutions to the max-flow problem were used to solve the
min-cut problem. These algorithms could be classified as those based on augmenting paths
\cite{NRMF97, NRMG97}, improvements to the augmenting path
approach based on blocking flows\cite{NRMH97, NRMI97} and those based on
pre-flow method introduced by Goldberg and Tarjan\cite{NRMJ97}.  The best known algorithms for the max-flow
problem are based on the preflow approach\cite{NRMP97, NRMQ97, NRMR97}.  The max-flow
problem also has been recently studied in a distributed setting in \cite{BCDK94}.

Further investigations revealed that there are more efficient direct
solutions to the min-cut problem (without solving max-flow and taking the dual).
Nagamochi and Ibaraki\cite{NRMK97} published the first deterministic
global minimum cut algorithm that is not based on flow, but was rather
complicated. Stoer and Wagner\cite{NRML97} presented a simple
deterministic global minimum cut algorithm which runs in $O(mn + n^{2}\log{n})$.

Karger\cite{NRMM97} presented the first randomized global min-cut
algorithm which runs in $O(mn^{2}\log^{3}n)$. The running time of a
single trial of the algorithm is $O(m\log^{2}{n})$. The algorithm has to
be repeated $n^{2}\log{n}$ times to achieve a high success probability
of $1-\frac{1}{n}$. Karger and Stein\cite{NRMN97} further improved its
running time to $O(n^{2}\log^{3}n)$ for the same probability.

Recently there has been revived interest in the min-cut problem
owing to its applications to network coding and wireless sensor networks
\cite{NRMS97, NRMT97, NRMU97}. Sensor networks operate in a
distributed setting and motivates a solution to the problem in a
distributed setting.

In this paper, we show how Karger's algorithm\cite{NRMM97} can be
adapted to efficiently solve the min-cut problem in a distributed
setting.  We assume a very general model of a graph where each node
knows only information about its neigbours. It is assumed that the
storage capacity of a node is bounded linearly in the size of the
number of its neigbours and  the computing capacity of a node is
bounded polynomially in the number of its neighbours. The assumption
is reasonable as each node must have storage and processing
capacity sufficient to keep track of communication with its neighbours.
The nodes can perform local computations and can communicate only with
its neighbours along the edges of the graph.  Our objective is to find
the value of the global min-cut and communicate the same to all the
nodes.  Moreover, each node must know which among the edges incident
on it are present in the min-cut computed. While the correctness proof follows directly from Karger's algorithm, the complexity analysis differs significantly. We show that  for a graph
of $n$ vertices and $m$ edges, the algorithm computes the global
min-cut with probability atleast $1-\frac{1}{n}$
with $O(mn^{2})$ message complexity and $O(n^{2})$ time complexity when
there is a global clock for synchronization.  We note that although
the assumption of a global clock may be impractical in applications
like sensor networks, there are standard techniques for converting
synchronous distributed algorithms to asynchronous algorithms, with
some loss in computational efficiency\cite{JHDS96}. We pursue the simpler synchronous setting here
as it allows a less cumbersome  presentation of the algorithm and a
simple analysis.

\section{The Algorithm}

\subsection{A Brief Description}
Assume that given a weighted graph $G = (V,E,w)$ where $E\subseteq V\times V$ and $w:E\rightarrow R^{+}\cup\{0\}$ is given(We use the terms network and graph interchangeably). In our algorithm $N_{u}$ represents the neighbourhood of vertex $u$, $weight_{u}$ represents the present edge weights of $N_{u}$, that is, for each $v\in N_{u}, weight_{u}[v]$ indicates the weight of edge $(u,v)$. $rank_{u}[v]$ is the rank of edge $(u,v)$, a random number which is uniformly chosen between 1 and $m^{k}$(for some fixed $k\geq 5$), on each trial. $maxrank$ represents the maximum value of rank among all the edges. Initially $maxrank_{u}$ is defined as the maximum rank of the edges connected to vertex $u$. The algorithm sets $maxrank = Max_{u\in V(G)} maxrank_{u}$. The $status$ of a vertex may be $ACTIVE$ or $INACTIVE$ (initially $ACTIVE$). $status_{u}=INACTIVE$ if all neighbouring edge weights of vertex $u$ are 0, which means that vertex cannot initiate the contraction process.  We call an edge {\em active} if at least one of its end points is active.

The algorithm proceeds by simulating edge contractions as in\cite{NRMM97}, by collecting vertices joined together by contraction into vertex {\em groups}. Edges within a group are inactive as they cannot be further contracted. At each step, an active edge of maximum rank is chosen for contraction. Since edge ranks are assigned uniformly at random, each active edge has equal probability for getting contracted. The algorithm continues contractions till only two vertex groups remain and the set of edges across the two groups is chosen as the mincut for that trial. The smallest cut found in $n^{2}\log{n}$ trials will be the mincut with probability $1-\frac{1}{n}$.

The variable $lastmsg_{u}$ stores the last message received at vertex $u$(used to reduce message flooding) and the boolean variable $stop_{u}$ is set to $true$ when only two vertex groups are remaining and no more contraction can be made, and set to $false$ otherwise.

The variable $g_{u}$ represents the present group id of vertex $u$, initially $g_{u}=u$. Initially there are $n$ groups, one for each vertex. As contractions progress, the number of groups reduces and we set $weight_{u}[v]=0$ if $g_{u}=g_{v}$ and $weight_{u}[v]\neq 0$ otherwise.  The following description presents a high level view of the algorithm.

\begin{algorithm}                      
\caption{distributed-mincut-in-a-nutshell()}   
\label{alg1}                           
\begin{algorithmic}
\STATE assign a $rank$ (between 1 and $m^{k}$) to each non-zero weighted edge. \COMMENT {Algorithm \ref{alg3}}
\STATE At each node $u$ of the network execute the following:
\STATE find $maxrank_{u}$ of each vertex $u$ locally. \COMMENT {Algorithm \ref{alg4}}
\STATE find the vertex $x$(with largest vertex id) having the maximum value of $maxrank$. \COMMENT {Algorithms \ref{alg5}, \ref{alg12}}
\IF[Algorithms \ref{alg6}, \ref{alg7}, \ref{alg13}, \ref{alg14}, \ref{alg20}]{there are only two groups}
\STATE compute local mincut $mc_{u}$ by summing the non-zero edge weights of vertex $u$. \COMMENT {Algorithm \ref{alg8}}
\STATE compute global mincut by summing up all local mincuts. \COMMENT {Algorithms \ref{alg9}, \ref{alg21}}
\STATE broadcast the mincut to all nodes and \textbf{stop}. \COMMENT {Algorithms \ref{alg10}, \ref{alg22}}
\ELSE
\STATE contract two vertex groups by making the edge weights between them zero and group ids equal to the value of $maxrank$ (The contraction process is initiated by the vertex $x$). \COMMENT {Algorithms \ref{alg6}, \ref{alg15}, \ref{alg16}}
\STATE repeat the algorithm
\ENDIF
\end{algorithmic}
\end{algorithm}

\subsection{Details of the Algorithm}

Each node in the network executes Algorithm \ref{alg2} described below. Here, the function $initialize()$ initializes the group id of each vertex with its vertex id. The function $assign$-$rank()$ assigns a $rank$ to each non-zero weighted edge with in the network, with a random value between 1 and $m^{k}$. The time complexity for this function is $O(n)$. The function $find$-$local$-$maxrank()$ computes the maximum rank within its neighbourhood, with time complexity $O(n)$. The function $find$-$global$-$maxrank()$ computes the maximum of all the $local$-$maxranks$ within the network, with time complexity $O(n)$ and message complexity $O(mn)$.

The function $check$-$eligibility$-$and$-$contract()$ checks whether there are more than two groups within the network and if so, contracts two groups by making all the edge weights between them zero and their group ids the same. This can be accomplished with time complexity $O(n)$ and message complexity $O(m)$. The function $check$-$termination$-$status()$ checks whether there are only two groups within the network and if so, invokes mincut computation and halts, otherwise the algorithm is repeated. This can be accomplished with time complexity $O(n)$ and message complexity $O(m)$. All the above mentioned functions except $initialize()$ has to be repeated $n-2$ times.

The function $find$-$local$-$mincut()$ computes the sum of edge weights within its neighbourhood, with time complexity $O(n)$. The function $find$-$global$-$mincut()$ computes the the sum of all $local$-$mincuts$ within the network, with time complexity $O(n^{2})$ and message complexity $O(mn)$.  Node $u$ messages to node $u+2^{i-1}$ in step $i$, for $i\in \{1,...\log{n}\}$ to ensure that the messages propagate to all nodes in $O(n^{2})$ time with only $O(mn)$ messages. The function $broadcast$-$mincut()$ broadcasts the computed mincut value to all the nodes within the network, which is done with time complexity $O(n)$ and message complexity $O(m)$. The function $synchronize()$ allows the nodes to wait for some time so that the same instruction can be executed by each node, in the next time step. This function waits for $O(n)$ steps.

\begin{algorithm}                      
\caption{distributed-mincut()\;\;\;\;\;//To be executed at each node}   
\label{alg2}                           
\begin{algorithmic}
\STATE initialize()
\REPEAT
\STATE assign-rank()
\STATE find-local-maxrank()
\STATE find-global-maxrank()
\STATE synchronize()
\STATE check-eligibility-and-contract()
\STATE synchronize()
\STATE check-termination-status()
\STATE synchronize()
\UNTIL{$stop_{u}=true$}
\STATE find-local-mincut()
\STATE find-global-mincut()
\STATE synchronize()
\STATE broadcast-mincut()
\end{algorithmic}
\end{algorithm}

\begin{algorithm}                      
\caption{initialize()}   
\label{alg25}                           
\begin{algorithmic}
\STATE $g_{u} \leftarrow u$
\end{algorithmic}
\end{algorithm}

\begin{algorithm}                      
\caption{assign-rank()}   
\label{alg3}                           
\begin{algorithmic}
\STATE \COMMENT {Rank of an edge to be assigned by higher numbered end-point}
\FOR{each $v\in N_{u}$}
\IF{$u>v$}
\IF{$weight_{u}[v]\neq 0$}
\STATE $rank_{u}[v] \leftarrow$ a random number between 1 and $m^{k}$
\ELSE
\STATE $rank_{u}[v] \leftarrow$ 0
\ENDIF
\STATE send(SET-RANK, $rank_{u}[v]$) to $v$. \COMMENT {See Algorithm \ref{alg11} for receipt of message}
\ENDIF
\ENDFOR
\end{algorithmic}
\end{algorithm}

\begin{algorithm}                      
\caption{find-local-maxrank()}   
\label{alg4}                           
\begin{algorithmic}
\STATE $maxrank_{u} \leftarrow max_{v\in N_{u}}(rank_{u}[v])$
\end{algorithmic}
\end{algorithm}

\begin{algorithm}                      
\caption{find-global-maxrank()}   
\label{alg5}                           
\begin{algorithmic}
\STATE send(FIND-MAX-RANK, $maxrank_{u}$) to each $v \in N_{u}$. \COMMENT {See Algorithm \ref{alg12} for receipt of message}
\end{algorithmic}
\end{algorithm}

\begin{algorithm}                      
\caption{check-eligibility-and-contract()}   
\label{alg6}                           
\begin{algorithmic}
\STATE $stop_{u}\leftarrow true$
\IF{$maxrank_{u}=max_{v\in N_{u}}(rank_{u}[v])$ and $u>v$}
\IF{$\exists w\in N_{u}$ with $weight_{u}[w]\neq 0$ and $v\neq w$ and $g_{v}\neq g_{w}$}
\STATE $stop_{u}\leftarrow false$
\STATE contract()
\ELSE
\STATE send(IS-ELIGIBLE-CONTRACT, $u$, $g_{u}$, $g_{v}$) to each $x\in N_{u}$. \COMMENT {See Algorithm \ref{alg13} for receipt of message}
\ENDIF
\ENDIF
\end{algorithmic}
\end{algorithm}

\begin{algorithm}                      
\caption{contract()}   
\label{alg15}                           
\begin{algorithmic}
\IF{$maxrank_{u}=max_{v\in N_{u}}(rank_{u}[v])$ and $u>v$}
\STATE $weight_{u}[v] \leftarrow 0$
\STATE check-active()
\STATE $g_{u}\leftarrow maxrank_{u}$
\STATE send(SET-GROUP-ID, $g_{u}$, $g_{v}$, $maxrank_{u}$) to each $x \in N_{u}$ with $weight_{u}[x]=0$. \COMMENT {See Algorithm \ref{alg16} for receipt of message}
\ENDIF
\end{algorithmic}
\end{algorithm}

\begin{algorithm}                      
\caption{check-termination-status()}   
\label{alg7}                           
\begin{algorithmic}
\STATE send(STOP, $stop_{u}$) to each $x\in N_{u}$. \COMMENT {See Algorithm \ref{alg20} for receipt of message}
\end{algorithmic}
\end{algorithm}

\begin{algorithm}                      
\caption{find-local-mincut()}   
\label{alg8}                           
\begin{algorithmic}
\IF{$status_{u}=ACTIVE$}
\STATE $mc_{u}\leftarrow \sum_{v\in N_{u}}{weight_{u}[v]}$
\ELSE
\STATE $mc_{u}\leftarrow 0$
\ENDIF
\end{algorithmic}
\end{algorithm}

\begin{algorithm}                      
\caption{find-global-mincut()}   
\label{alg9}                           
\begin{algorithmic}
\FOR{$i\leftarrow 1$ to $\log{n}$ step by 1}
\FOR{$j\leftarrow 2^{i-1}$ to $n-1$ step by $2^{i}$}
\IF{$u=j$}
\STATE send(LOCAL-MC, $mc_{u}$, $u$, min($u+2^{i-1}$, $n$)) to each $v\in N_{u}$. \COMMENT {See Algorithm \ref{alg21} for receipt of message}
\ENDIF
\STATE synchronize()
\ENDFOR
\ENDFOR
\end{algorithmic}
\end{algorithm}

\begin{algorithm}                      
\caption{broadcast-mincut()}   
\label{alg10}                           
\begin{algorithmic}
\IF{$u=n$}
\STATE $mc_{u}\leftarrow mc_{u} / 2$
\STATE send(MINCUT, $mc_{u}$, $u$) to each $v\in N_{u}$. \COMMENT {See Algorithm \ref{alg22} for receipt of message}
\ENDIF
\end{algorithmic}
\end{algorithm}

\begin{algorithm}                      
\caption{upon receipt of (SET-RANK, $num$) msg from $w$}   
\label{alg11}                           
\begin{algorithmic}
\STATE $rank_{u}[w]\leftarrow num$
\end{algorithmic}
\end{algorithm}

\begin{algorithm}                      
\caption{upon receipt of (FIND-MAX-RANK, $m$) msg from $w$}   
\label{alg12}                           
\begin{algorithmic}
\STATE\COMMENT{find maximum rank among all vertices}
\IF{$m>maxrank_{u}$}
\STATE $maxrank_{u}\leftarrow m$
\STATE send(FIND-MAX-RANK, $m$) to each $v \in N_{u}$ where $v\neq w$
\ENDIF
\end{algorithmic}
\end{algorithm}

\begin{algorithm}                      
\caption{upon receipt of (IS-ELIGIBLE-CONTRACT, $v$, $g'$, $g''$) msg from $w$}   
\label{alg13}                           
\begin{algorithmic}
\STATE\COMMENT{checks the eligibility of contraction}
\IF{(IS-ELIGIBLE-CONTRACT, $v$, $g'$, $g''$) $\neq lastmsg_{u}$}
\IF{$\exists y\in N_{u}$ with $weight_{u}[y]\neq 0$ and $g_{y}\neq g'$ and $g_{y}\neq g''$}
\STATE send(ELIGIBLE-CONTRACT, $v$, $g'$) to each $z\in N_{u}$ with $weight_{u}[z]=0$ or ($weight_{u}[z]\neq 0$ and $g_{z}=g'$). \COMMENT {See Algorithm \ref{alg14} for receipt of message}
\ELSE
\STATE send(IS-ELIGIBLE-CONTRACT, $v$, $g'$, $g''$) to each $z\in N_{u}$ with $weight_{u}[z]=0$ and $z\neq w$
\ENDIF
\STATE $lastmsg_{u}\leftarrow$ (IS-ELIGIBLE-CONTRACT, $v$, $g'$, $g''$)
\ENDIF
\end{algorithmic}
\end{algorithm}

\begin{algorithm}                      
\caption{upon receipt of (ELIGIBLE-CONTRACT, $v$, $g'$) msg from $w$}   
\label{alg14}                           
\begin{algorithmic}
\IF{(ELIGIBLE-CONTRACT, $v$, $g'$) $\neq lastmsg_{u}$}
\IF{$u=v$}
\STATE $stop_{u}\leftarrow false$
\STATE contract()
\ELSE
\STATE send(ELIGIBLE-CONTRACT, $v$, $g'$) to each $z\in N_{u}$, $z\neq w$ with $weight_{u}[z]=0$ or ($weight_{u}[z]\neq 0$ and $g_{z}=g'$)
\ENDIF
\STATE $lastmsg_{u}\leftarrow$ (ELIGIBLE-CONTRACT, $v$, $g'$)
\ENDIF
\end{algorithmic}
\end{algorithm}

\begin{algorithm}                      
\caption{upon receipt of (SET-GROUP-ID, $g'$, $g''$, $newrank$) msg from $w$}   
\label{alg16}                           
\begin{algorithmic}
\STATE\COMMENT {update group id of all vertices in the groups g' and g'' by $maxrank_{u}$ by sending messages}
\IF{$g_{u}\neq newrank$}
\STATE $weight_{u}[w]\leftarrow 0$
\STATE check-active()
\STATE $g_{u}\leftarrow newrank$
\IF{$status_{u}=ACTIVE$}
\FOR{all $v \in N_{u}$ with $weight_{u}[v]\neq 0$}
\IF{$g_{v}=g'$ or $g_{v}=g''$ or $g_{v}=newrank$}
\STATE $weight_{u}[v]\leftarrow 0$
\STATE check-active()
\STATE send(SET-WEIGHT) to $v$. \COMMENT {See Algorithm \ref{alg18} for receipt of message}
\ENDIF
\ENDFOR
\ENDIF
\STATE send(SET-GROUP-ID, $g'$, $g''$, $newrank$) to each $x \in N_{u}$ where $weight_{u}[x]=0$
\ENDIF
\end{algorithmic}
\end{algorithm}

\begin{algorithm}                      
\caption{synchronize()}   
\label{alg17}                           
\begin{algorithmic}
\STATE\COMMENT {waits for all nodes to reach the same step of algorithm}
\STATE wait for $n$ pulses
\end{algorithmic}
\end{algorithm}

\newpage

\begin{algorithm}                      
\caption{upon receipt of (SET-WEIGHT) msg from $w$}   
\label{alg18}                           
\begin{algorithmic}
\STATE $weight_{u}[w]\leftarrow 0$
\STATE check-active()
\end{algorithmic}
\end{algorithm}

\begin{algorithm}                      
\caption{check-active()}   
\label{alg19}                           
\begin{algorithmic}
\IF{$\forall v\in N_{u}, weight_{u}[v]=0$}
\STATE $status_{u}=INACTIVE$
\ENDIF
\end{algorithmic}
\end{algorithm}

\begin{algorithm}                      
\caption{upon receipt of (STOP, $t$) msg from $w$}   
\label{alg20}                           
\begin{algorithmic}
\STATE\COMMENT {broadcast the information on the number of groups in the network}
\IF{(STOP, $t$)$\neq lastmsg_{u}$}
\IF{$t=false$}
\STATE $stop_{u}\leftarrow false$
\STATE send(STOP, $t$) to each $x\in N_{u}$
\STATE $lastmsg_{u}\leftarrow$ (STOP, $t$)
\ENDIF
\ENDIF
\end{algorithmic}
\end{algorithm}

\begin{algorithm}                      
\caption{upon receipt of (LOCAL-MC, $mcut$, $x$, $v$) msg from $w$}   
\label{alg21}                           
\begin{algorithmic}
\STATE\COMMENT {computes mincut partially}
\IF{(LOCAL-MC, $mcut$, $x$, $v$)$\neq lastmsg_{u}$}
\IF{$u=v$}
\STATE $mc_{u}\leftarrow mc_{u} + mcut$
\ELSE
\STATE send(LOCAL-MC, $mcut$, $x$, $v$) to each $y\in N_{u}$
\ENDIF
\STATE $lastmsg_{u}\leftarrow$ (LOCAL-MC, $mcut$, $x$, $v$)
\ENDIF
\end{algorithmic}
\end{algorithm}

\begin{algorithm}                      
\caption{upon receipt of (MINCUT, $v$, $mincut$) msg from $w$}   
\label{alg22}                           
\begin{algorithmic}
\STATE\COMMENT {broadcasts the mincut to all nodes}
\IF{(MINCUT, $v$, $mincut$)$\neq lastmsg_{u}$}
\STATE $mc_{u}\leftarrow mincut$
\STATE send(MINCUT, $v$, $mincut$) to each $y\in N_{u}$
\STATE $lastmsg_{u}\leftarrow$ (MINCUT, $v$, $mincut$)
\ENDIF
\end{algorithmic}
\end{algorithm}

\subsection{Correctness}
First, we bound the probability of error created by edges getting the same rank.

\newtheorem{mydef12}{Lemma}[subsection]
\begin{mydef12}
 The probability that two edges get the same rank in $n$ trials is $O(n^{-2})$.
\end{mydef12}

\begin{proof}
The rank is a value from the set $\{1...m^{k}\}$. The probability that two edges $m$ and $m'$ having the same rank,
$Pr[rank(m)=rank(m')] \leq \frac{1}{m^{k}}$
\\Hence, $Pr[\exists (m,m'): rank(m)=rank(m')] \leq \\ \sum_{(m,m')\in E\times E}{Pr[rank(m)=rank(m')]} \leq \frac{m^{2}}{m^{k}} = \frac{1}{m^{k-2}}$
\newline
Thus, using the union bound, probability that there exists two edges $m$ and $m'$ having the same rank in $n$ iterations is
$\leq \frac{n}{m^{k-2}} \leq \frac{m}{m^{k-2}} = \frac{1}{m^{k-3}}$.
Now choose $k \geq 5$. Then, $Pr[rank(m)=rank(m')] \leq \frac{1}{m^{2}} = O(n^{-2})$.
\end{proof}

The following Lemma proceeds exactly as in \cite{NRMM97}.
\newtheorem{mydef13}[mydef12]{Lemma}
\begin{mydef13}
 A particular min-cut in G is produced by the contraction algorithm with probability $\Omega(n^{-2})$.
\end{mydef13}

\begin{proof}
Let $c$ be the value of the mincut in $G$. Each contraction reduces the number of vertices in the graph by one. Consider the contraction executed when the graph has $r$ vertices. Since the contracted graph has a min-cut of at least $c$, it must have minimum degree $c$, and thus atleast $\frac{rc}{2}$ edges. However, only $c$ of these edges are in min-cut. Thus, a randomly chosen edge is in the min-cut with probability at most $\frac{2}{r}$. The probability that we never contract a min-cut edge through all $n-2$ contractions is atleast
$(1-\frac{2}{n})(1-\frac{2}{n-1})(1-\frac{2}{n-2})....(1-\frac{2}{3}) = \binom{n}{2}^{-1} = \Omega(n^{-2})$
\end{proof}

\newpage

\subsection{Complexity Analysis}
\subsubsection{Message complexity}
\newtheorem{mydef14}{Theorem}[subsection]
\begin{mydef14}
 The Karger's distributed algorithm uses $O(mn^{2})$ messages, in a single trial.
\end{mydef14}

\begin{proof}
It is not hard to see that the most expensive steps in a trial are those of determination of $maxrank$ from local maxranks(find-global-maxrank()) and that of computing the mincut at the end(find-global-mincut()). In find-global-maxrank(), each node sends its local maxrank value to its neighbours and this is repeated atmost $n$ times(number of times equal to the diameter of the graph sufficies). Hence the total number of messages is bounded by $nO(m+n) = O(mn)$. Thus the message complexity for $n-2$ iterations per trial is $O(mn^{2})$.  Finally, in step $i$ of find-global-mincut(), $\frac{n}{2^{i}}$ nodes send messages to its neighbours. The total number of messages sent at each step is bounded by $O(m)$. Thus, the total number of messages is $\Sigma_{i=1}^{\log{n}} \frac{nm}{2^{i}} = O(mn)$.  Hence the overall message complexity is $O(mn^{2}) + O(mn) = O(mn^{2})$.
\end{proof}

\subsubsection{Time complexity}
\newtheorem{mydef15}[mydef14]{Theorem}
\begin{mydef15}
The Karger's distributed algorithm computes mincut in $O(n^{2})$ time, in a single trial.
\end{mydef15}

\begin{proof}
Before contraction, the algorithm assigns a rank (random number) to each edge and finds the max-rank among all the vertices in the graph. This requires atmost $n-1$ steps(strictly, number of steps equal to the diameter of the graph). For contraction, a message is sent from a vertex within one group to other group and the message is propagated to all the vertices within the second group and the neighbouring vertices of that group, which also takes atmost $n-1$ pulses. Since only one contraction can take place at any time and there are $n-2$ such contractions, the running time is $O(n^{2})$.  To estimate time for computing the mincut, the function find-global-mincut() runs $O(\log{n})$ steps and in step $i$, $\frac{n}{2^{i}}$ nodes flood the network. Thus the time per step is  $\frac{n^{2}}{2^{i}}$. Hence the total complexity is $\Sigma_{i=1}^{\log{n}} \frac{n^{2}}{2^{i}} = O(n^{2})$.
\end{proof}

\section{Conclusion and Future work}
A synchronous distributed version of the Karger's randomized algorithm under network setting is presented in this paper with a proof of correctness and complexity analysis. The present algorithm appears not to make use of the full power of parallelism available. It is interesting to look at how to efficiently reduce time and message complexity by conducting edge contractions in parallel.

%
\bibliographystyle{abbrv}
\bibliography{bibname}  
%
%
\end{document}